# An Investigation of Minimum Data Requirement for Successful Structure Determination of Pf2048.1 with REDCRAFT

Casey A. Cole[1], Daniela Ishimaru[2], Mirko Hennig[3], and Homayoun Valafar[1*]

[1]Department of Computer Science and Engineering, University of South Carolina, Columbia, SC 29208, USA

[2]Department of Biochemistry and Molecular Biology, Medical University of South Carolina, Charleston, SC 29425 USA

[3]Nutrition Research Institute, University of North Carolina at Chapel Hill, Kannapolis, NC 27514, USA

[*] Corresponding Author Email: homayoun@cec.sc.edu Phone: 1 803 777 2404 Fax: 1 803 777 3767

Mailing Address: Swearingen Engineering Center, Department of Computer Science and Engineering, University of South Carolina, Columbia, SC 29208, USA

**Abstract** – Traditional approaches to elucidation of protein structures by NMR spectroscopy rely on distance restraints also know as nuclear Overhauser effects (NOEs). The use of NOEs as the primary source of structure determination by NMR spectroscopy is time consuming and expensive. Residual Dipolar Couplings (RDCs) have become an alternate approach for structure calculation by NMR spectroscopy. In this work we report our results for structure calculation of the novel protein PF2048.1 from RDC data and establish the minimum data requirement for successful structure calculation using the software package REDCRAFT. Our investigations start with utilizing four sets of synthetic RDC data in two alignment media and proceed by reducing the RDC data to the final limit of {CN, NH} and {NH} from two alignment media respectively. Our results indicate that structure elucidation of this protein is possible with as little as {CN, NH} and {NH} to within 0.533Å of the target structure.

## 1 Introduction

Proteins are a class of organic macromolecules that perform many important biochemical functions in biological cells. Protein functions run the entire gamut from structural support and transport of biomaterial, to performing important enzymatic activities within a living organism. Unlike the genetic material (DNA/RNA) within Eukaryotic cells, cytosolic proteins are not protected with an additional bilayer membrane of the nucleus. Therefore, design and delivery of protein-based intervention of diseases is more pragmatic in the near future than genetic treatment of diseases. Furthermore, principles of modern biology stress the importance of protein structure and its function. Therefore, knowledge of protein structures becomes paramount in understanding the mechanism of their function (or dysfunction) and subsequently, intelligent and appropriate drug design.

An understanding of protein structure at atomic resolution serves as the first and critical step in understanding the molecular basis of nearly all diseases. While structure determination of proteins is becoming more routine, the cost of structure determination remains the prohibitive factor. Thanks to improvements by the Structural Genomics Initiative[1], [2] and Protein Structure Initiative[3], the cost of experimental structure determination of proteins has been reduced from approximately $1,000,000 per protein to $100,000. Although this is a significant reduction in cost, it is still an impediment in achieving personalized medicine where nearly 100,000 protein structures will need to be determined for each person. This approximate cost of $10^{10}$ per person clearly represents a significant economical barrier.

In recent years, the use of Residual Dipolar Coupling (RDC) data acquired from Nuclear Magnetic Resonance (NMR) spectroscopy has become a potential avenue for a significant reduction in the cost of structure determination of proteins. Recent work[4]–[7] has demonstrated the challenges in structure calculation of proteins from RDC data alone, and some potential solutions have been introduced[5], [6], [8]. One such approach named REDCRAFT[4], [9], [10] has been demonstrated to be successful in structure calculation of proteins from a reduced set of RDC data. The main objective in this research is to perform a feasibility study for structure calculation of a novel protein from RDC data. Our feasibility study will establish the minimum required data for unambiguous structure calculation that is optimized for a given protein. A better understanding of minimum data requirement will help to alleviate the cost of structure determination by avoiding acquisition of unneeded data. To accomplish this objective we use a suggested structure of PF2048.1 as an approximate template for its native structure. Albeit it is clear that the suggested structure is not the native structure, we have mounting evidence that the native structure is less than 4Å away. We argue that our findings from a suggested structure is relevant to the actual structure due to their close structural resemblance.

## 2 Background and Method

### 2.1 Residual Dipolar Couplings

RDCs can be acquired via NMR spectroscopy. The theoretical basis of RDC interaction had been established and experimentally observed in 1963 [11]. However, it has only become a more prevalent source of data for structure determination of biological macromolecules in recent years due to availability of alignment media. Upon the reintroduction of order to an isotropically tumbling molecule, RDCs can be easily acquired. The RDC interaction between two atoms in space can be formulated as shown in Eq. (1).

$$D_{ij} = D_{max} \left\langle \frac{3\cos^2(\theta_{ij}(t)) - 1}{2} \right\rangle \quad (1)$$

$$D_{max} = \frac{-\mu_0 \gamma_i \gamma_j h}{(2\pi r)^3} \quad (2)$$

In this equation, $D_{ij}$ denotes the residual dipolar coupling in units of Hz between nuclei $i$ and $j$. The $\theta_{ij}$ represents the time-dependent angle of the internuclear vector between nuclei $i$ and $j$ with respect to the external magnetic field, and the angle brackets signify time averaging. In Eq. (2), $D_{max}$ represents a scalar multiplier dependent on the two interacting nuclei. In this equation, $\gamma_i$ and $\gamma_j$ are nuclear gyromagnetic ratios, $r$ is the internuclear distance (assumed fixed for directly bonded atoms), $h$ is the modified Planck's constant and $\mu_0$ represents the permeability of free space.

### 2.2 REDCRAFT Structural Fitness Calculation

While generating a protein structure from a given set of residual dipolar couplings is nontrivial, it is straightforward to determine how well a given structure fits a set of RDCs. Through algebraic manipulation of Eq. (1) RDC interaction can be represented as shown in Eq. (3),

$$D_{ij} = v_{ij} * S * v_{ij}^T \quad (3)$$

where $S$ represents the Saupe order tensor matrix [11] and $v_{ij}$ denotes the normalized interacting vector between the two interacting nuclei $i$ and $j$. REDCRAFT takes advantage of this principle by quantifying the fitness of a protein to a given set of RDCs (in units of Hz) and calculating a root-mean-squared deviation as shown in Eq. (4). In this equation $D_{ij}$ and $D'_{ij}$ denote the computed and experimentally acquired RDCs respectively, $N$, represents the total number of RDCs for the entire protein, and $M$ represents the total number of alignment media in which RDC data have been acquired. In this case a smaller fitness value indicates a better structure.

$$Fitness = \sqrt{\frac{\sum_{j=1}^{M} \sum_{i=1}^{N} (D_{ij} - D'_{ij})^2}{M * N}} \quad (4)$$

The REDCRAFT algorithm and its success in protein structure elucidation has been previously described and documented in detail [4], [9], [10], [12], [13]. Here we present a brief overview. REDCRAFT calculates structures from RDCs using two separate stages. In the first stage (*Stage-I*), a list of all possible discretized torsion angles is created for each pair of adjoining peptide planes. This list is then filtered based on allowable regions within the Ramachandran space [14]. The list of torsion angles that remain are then ranked based on fitness to the RDC data. These lists of potential angle configurations are used to reduce the search space for the second stage.

*Stage-II* begins by constructing the first two peptide planes of the protein. Every possible combination of angles from *Stage-I* between peptide planes $i$ and $i+1$ are evaluated for fitness with respect to the collected data, and the best $n$ candidate structures are selected, where $n$ denotes the search depth. The list of dihedral angles corresponding to the top $n$ structures are then combined with every possible set of dihedral angles connecting the next peptide plane to the current fragment. Each of these candidate structures is evaluated for fitness and the best $n$ are again selected and carried forward for additional rounds of elongation. All combination of dihedral angles worse than the best $n$ are eliminated, thus removing an exponential number of candidate structures from the search space. This elongation process is repeated iteratively, incrementally adding peptide planes until the entire protein is constructed.

The number of RDCs required to correctly fold a novel protein with a bundle of four nearly parallel helices with REDCRAFT has not been previously examined in a systematic manner. Here we investigate the effect of reducing the available RDCs on the quality of the resulting computational structure. Collecting fewer RDCs per peptide plane can substantially reduce data collection times. In particular, $^{15}N$-$^1H$ RDCs are easily collected because they avoid expensive $^{13}C$ labeling. Furthermore, $^{15}N$-$^1H$ RDC values are typically large in magnitude, reducing the effect of measurement error. $C_\alpha$-$H_\alpha$ RDCs are large in magnitude but require $^{13}C$ labeling, complicating sample preparation. RDCs for additional vectors can be collected, but with a decreasing utility and at a greater expense.

## 2.3 PF2048.1 Protein

The novel protein PF2048.1 is a 9.16 kDa, 71 residue monomeric protein with less than 17% sequence identity to any structurally characterized protein in PDB (as of April, 2015) serves as the primary target of our investigations. PF2048.1 was expressed in *E. coli* as an N-terminal $His_6$-GB1 fusion that can be efficiently cleaved by TEV protease introducing a single (non-native) Gly residue at position –1. Nearly complete assignments for backbone and sidechain protons, carbons and nitrogens were obtained using standard methods. The resulting 1045 NOE restraints together with TALOS backbone torsion restraints were employed to determine an experimental target structure. Using this reference structure and the structural alignment software 3D-Blast [15] we were able to investigate the structural uniqueness of PF2048.1. Of the resulting proteins 1AEP, a 161 residue apolipoprotein, was identified as the top entry with the highest 3D-Blast score (score of 54.4). We then utilized msTALI [16] to align 1AEP and PF2048.1 based on structural similarity. The final alignment identified 26 residues to be structurally conserved to within 2.9Å between the two proteins, corresponding to about 36% (26 conserved / 71 total residues = 0.36) structural similarity. Figure 1 shows the resulting alignment between the two structures. The two's overall structural deviation was calculated to be 5.265Å.

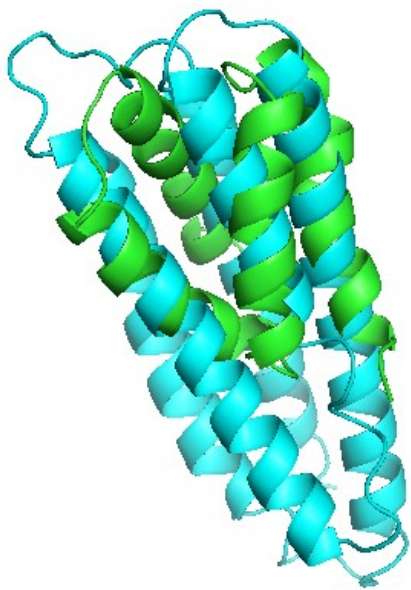

Figure 1. NOE structure of PF2048.1 (green) aligned to 1AEP (blue) using PyMOL. According to PyMOL the two exhibited structural dissimilarity of 5.265Å.

Due to its novelty in both sequence and structure PF2048.1 is an ideal candidate to study the effectiveness of computing protein structure from solely residual dipolar couplings. In addition, the unique arrangement of the helical secondary structural elements of this protein will provide a realistic exploration of the challenges that REDCRAFT will be faced during structure calculation purely from RDCs.

## 2.4 Simulated RDC Data

Using REDCAT [17], [18] and the reference structure residual dipolar couplings were simulated in two alignment media using the order tensors in Table 1. Error of ±1Hz was uniformly added across all N-H vectors to simulate mild experimental noise in the data sets. Similarly, RDC data from other vectors were distorted by uniformly distributed noise in a range proportional to the expected range of RDCs. These level of random noise for each vector type is shown in Table 2. In addition, Table 2 summarizes the minimum and maximum values corresponding to these order tensors for each RDC vector.

Table 1. Order tensors used for synthetic RDC calculation.

|    | Sxx | Syy | Szz | Alpha | Beta | Gamma |
|----|-----|-----|-----|-------|------|-------|
| M1 | $3 \times 10^{-4}$ | $5 \times 10^{-4}$ | $-8 \times 10^{-4}$ | 0 | 0 | 0 |
| M2 | $-4 \times 10^{-4}$ | $-6 \times 10^{-4}$ | $10 \times 10^{-4}$ | 40 | 50 | -60 |

Table 2. Columns 2 and 3 display minimum and maximum RDC values for each vector set using the order tensors in Table 1 in two alignment media (M1 and M2). The last column summarizes the range of uniformly distributed noise that was added to each dataset.

|    | RDC | Minimum | Maximum | Added noise |
|----|-----|---------|---------|-------------|
| M1 | N-C | -2.029 | 1.287 | ±0.1Hz |
|    | N-H | -18.904 | 11.815 | ±1Hz |
|    | C-H | -3.557 | 5.692 | ±0.3Hz |
|    | $C_\alpha$-$H_\alpha$ | -23.32 | 37.312 | ±1.97Hz |
| M2 | N-C | -1.544 | 2.574 | ±0.1Hz |
|    | N-H | -14.178 | 23.63 | ±1Hz |
|    | C-H | -7.115 | 4.269 | ±0.3Hz |
|    | $C_\alpha$-$H_\alpha$ | -46.64 | 27.984 | ±1.97Hz |

## 2.5 Evaluation

Our evaluation will proceed by incremental reduction in the data quantity; maintaining the RDC data that are easiest to acquire from NMR spectroscopy. To that end, we will proceed by first eliminating $C_\alpha$-$H_\alpha$ RDC data from both alignment media since its acquisition increases the cost of protein production significantly. The second phase of our investigation will focus on reducing the RDC data sets from 3 RDCs per alignment medium, to 3 from medium 1 and 1 from medium 2, followed by 2 from medium 1 and 1 from medium 2.

The software REDCRAFT will be utilized for our structure calculation without refinement in any other auxiliary program such as Xplor-NIH[19] or CNS[20]. We anticipated that consistent with principles of Information Theory, more extensive search parameters of REDCRAFT will need to be enabled as a function of reduced datasets to compensate for the absence of information.

The software package PyMOL[21] was utilized in order to calculate the bb-rmsd (backbone root mean squared deviation) between the REDCRAFT structure and the target structure (the NOE structure from which the RDC data were generated). The measure of bb-rmsd is prevalently used to establish the structure similarity between two proteins and values under 3.5Å can signify presence of structural resemblance, while values under 2Å can be interpreted as strong structural resemblance. Our objective is to calculate structures of PF2048.1 using REDCRAFT that exhibit structural similarity to the target protein under 2Å.

The other measure we will use to evaluate structures is the RDC fitness score calculated by REDCRAFT (discussed in detail in section 2.2). This fitness score provides information about how well the RDCs fit the final structure. A score is considered to be of high quality if its score falls at or below the error level of the data (in our case <1Hz). The lower the score the better the structure.

## 3 Results and Discussion

To evaluate the ability of REDCRAFT to predict the correct structure of PF2048.1, five test cases were established. In each of the cases the amount of data was varied to simulate five different possible data sets. The data sets are summarized in Table 3.

Table 3. Summary of the RDCs used in each experiment.

| Set | Medium # | RDCs Utilized |
|---|---|---|
| 1 (4,4) | 1 | {C-N, N-H, C-H, $C_\alpha$-$H_\alpha$} |
| | 2 | {C-N, N-H, C-H, $C_\alpha$-$H_\alpha$} |
| 2 (4,1) | 1 | {C-N, N-H, C-H, $C_\alpha$-$H_\alpha$} |
| | 2 | {N-H} |
| 3 (3,3) | 1 | {C-N, N-H, C-H} |
| | 2 | {C-N, N-H, C-H} |
| 4 (3,1) | 1 | {C-N, N-H, C-H} |
| | 2 | {N-H} |
| 5 (2,1) | 1 | {C-N, N-H] |
| | 2 | {N-H} |

In the sections that follow we will report our findings in each of the cases in Table 3 to evaluate the feasibility of successful protein structure elucidation with the given data set.

### 3.1 Structure calculation from 4, 4 RDCs

In this experiment the following RDCs corresponding to the vector set {CN, NH, CH, $C_\alpha H_\alpha$} were utilized in two alignment media. The configuration of REDCRAFT is summarized in Table 4 below:

Table 4. Parameters of REDCRAFT for experiment 1 where in the Decimation Parameters C.S. denotes Cluster Sensitivity and S.T. denotes Score Threshold.

| Search Depth | Decimation Parameters | | Minimization | Lennard Jones Cutoff |
|---|---|---|---|---|
| 200 | C.S. | S.T. | Performed every residue | 50.0 |
| | 4 | 1.0 | | |

The resulting structure, seen in Figure 2 was measured to have a REDCRAFT fitness score of 0.776 and showed 1.035Å of structural deviation from our target structure.

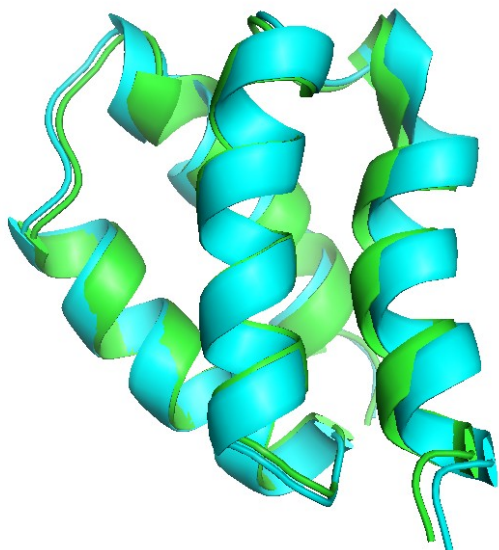

Figure 2. Resulting structure (in green) superimposed to the target target structure (in blue). The two exhibited structural difference of 1.035Å.

### 3.2 Structure calculation from 4,1 RDCs

In this experiment two different sets of RDC data were used in both alignment media. The first set contained four vectors {CN, NH, CH, $C_\alpha H_\alpha$} and the second just one vector set {NH}. The corresponding REDCRAFT parameters for this exercise are summarized in Table 5. Consistent with our expectation, due to the reduction in data quantity, a more thorough search by REDCRAFT was required in order to achieve a comparable result to that of the (4,4) exercise. The more thorough search was achieved through the adjustment of the C.S. an S.T. terms. The adjustment of these two terms allow for a more refined clustering of the search space as a function of reduced dataset *N* in Eq. (4).

Table 5. Parameters of REDCRAFT for experiment 2.

| Search Depth | Decimation Parameters | | Minimization | Lennard Jones Cutoff |
|---|---|---|---|---|
| 200 | C.S. | S.T. | Performed every 3rd residue | 50.0 |
| | 3 | 0.8 | | |

The resulting structure (seen in Figure 3) exhibited a RDC fitness score of 0.741 and a bb-rmsd of 1.594Å with respect to the target structure.

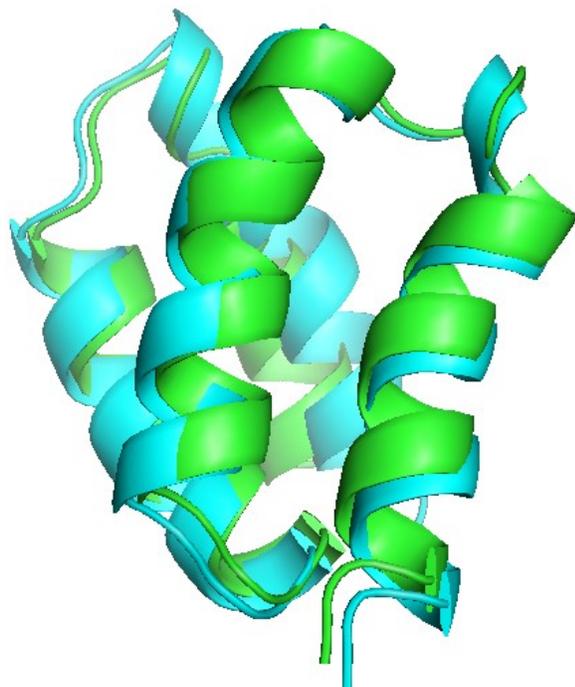

Figure 3. Resulting structure (in green) superimposed to the target structure (in blue). The two showed structural deviation of 1.594Å.

### 3.3 Structure calculation from 3,3 RDCs

In this experiment two sets of three RDCs {CN, NH, CH} were utilized. Several REDCRAFT configurations (similar to those in experiment 1 and 2) were attempted on this dataset but it became clear that there was something inherently anomalous about constructing a protein with these two particular sets of RDCs. As a result of these difficulties we were forced to incorporate additional secondary structural information and perform a more directed folding process. In our case the phi and psi angles were restricted to oscillate in the range of [-60:-50] and [-50:-40] respectively for the helical residues 3-16, 22-35, 39-52 and 57-70. The addition of secondary structural constraints can easily be facilitated through the use of secondary structure prediction tools such as Jpred, Jpred3 and I-TASSER[22]–[24], or through early interpretation of the data available from NMR spectroscopy without imposing any additional data acquisition costs.

The resulting structure (seen in Figure 4) had a RDC fitness score of 0.382 and a bb-rmsd of 1.002Å with respect to the target structure.

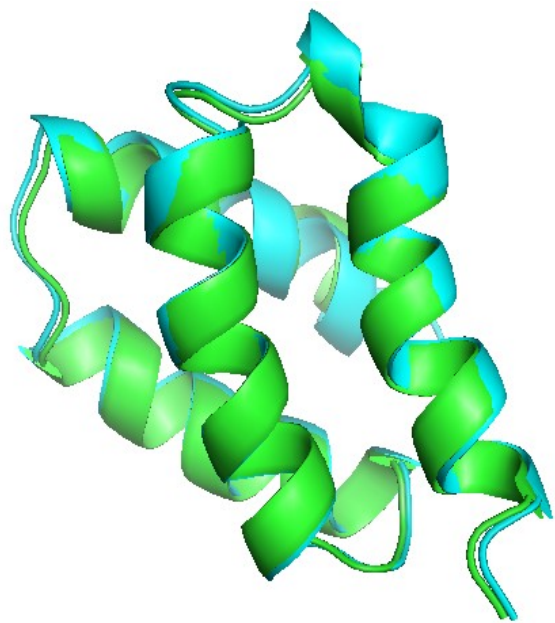

Figure 4. Resulting structure (in green) aligned to the target structure (in blue). The two showed structural deviation of just 1.001Å.

### 3.4 Structure calculation from 3,1 RDCs

In this experiment two different sets of RDCs were used; the first set containing three vectors {CN, NH, CH} and the second containing just one vector {NH}. The REDCRAFT configuration is summarized in the Table 6 below:

Table 6. Parameters of REDCRAFT for experiment 4.

| Search Depth | Decimation Parameters | | Minimization | Lennard Jones Cutoff |
|---|---|---|---|---|
| 200 | C.S. | S.T. | Performed every 3rd residue | 50.0 |
| | 3 | 1.0 | | |

Surprisingly, this combination of data (although a subset of the 3,3 exercise) was less refractory and did not require the incorporation of dihedral restraints or modification of search parameters in order to perform a more extensive search of the solution space. The resulting structure, as seen in Figure 5, exhibits a RDC fitness score of 0.741 and bb-rmsd from the target structure of 1.594Å, mirroring the results in 3.2.

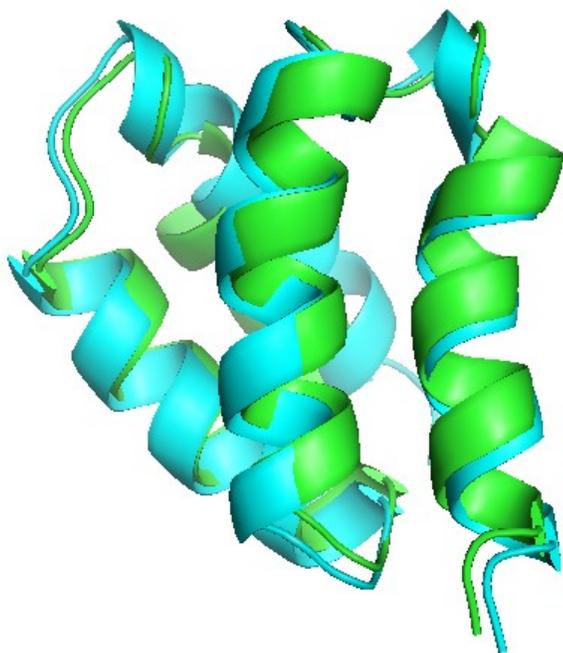

Figure 5. Resulting structure (in green) superimposed to the target structure (in blue). As in experiment 2, the two exhibited a backbone RMSD of 1.594Å.

### 3.5 Structure calculation from 2,1 RDCs

The final experiment in establishing the boundaries of data requirement is based on {CN, NH} and {NH}. Due to further reduction of the datasets we were again forced to incorporate secondary structure constraints along with the following REDCRAFT parameters summarized in Table 7. The ranges for the secondary structure constraints remained the same as that of the experiment described in 3.3.

Table 7. REDCRAFT parameters for experiment 5 utilizing 2,1 RDC sets resulting in a structure 3.03Å from the target structure.

| Search Depth | Decimation Parameters | | Minimization | Lennard Jones Cutoff |
|---|---|---|---|---|
| 200 | C.S. | S.T. | Performed every residue | 50.0 |
| | 3 | 0.5 | | |

The resulting structure in this experiment showed structural deviation from the target structure of 3.03Å—a bb-rmsd that indicates need for further refinement. Careful investigation of the changes in RDC fitness scores revealed that midway through the last helix (around residue 64) there was a significant spike in fitness to RDC data (as seen in Figure 6). This prompted a fragmented study of this protein where the structure is determined in two contiguous segments. Since the spike occurred in the middle of a helix, we chose to terminate the first segment at residue 57 (the beginning of the affected helix) in an attempt to conserve secondary structure elements as much as possible. This approach yielded two fragments [1:56] and [57:72] having bb-rmsd's to the target structure of 0.465Å and 0.724Å respectively. Using RDCs to predict the orientation of the two fragments (as previously shown in theory [25]) we properly oriented and connected the two fragments. The resulting structure (seen in Figure 7) exhibited a RDC fitness score of 0.173 and bb-rmsd of 0.533Å to the target structure.

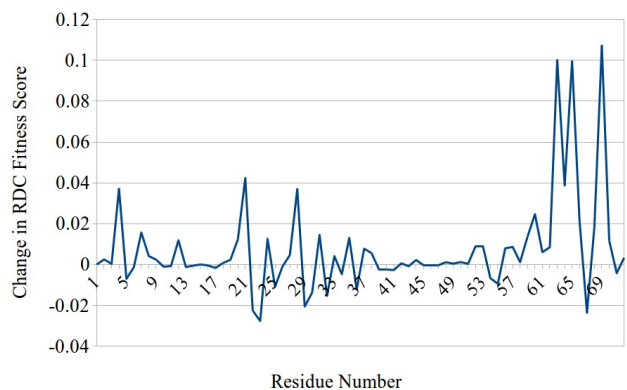

Figure 6. Graph showing the change in RDC fitness (y-axis) throughout the 72 residues (x-axis). A spike can be seen to occur at residue 64.

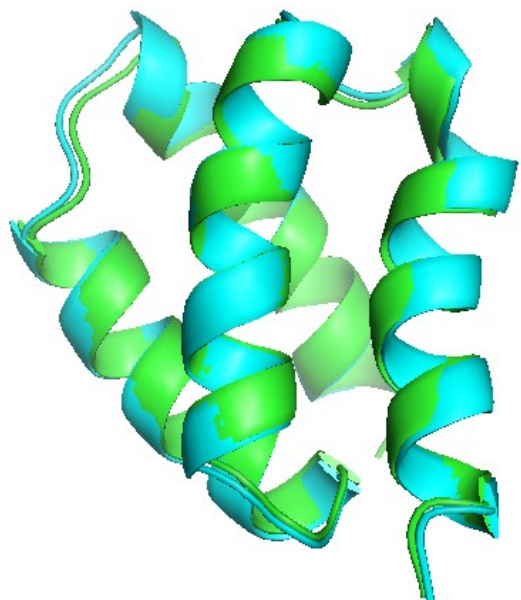

Figure 7. Resulting structure (green) from experiment 5 superimposed to the target structure (blue). The structural deviation between the two was calculated to be 0.533Å.

## 4 Conclusion

Exploration of the minimum data requirement is useful in order to establish the expected financial cost of a protein's structure determination. An exploration mechanism such as the one presented here will allow for appropriate allocation of funds as a function of a protein's medical or biological importance. This is a critical contribution to the repertoire of structure determination approaches especially in the context of personalized medicine where funds can be appropriate allocated toward culprit proteins in human diseases.

Our investigation through exploration of the five exercises listed in the previous section, has revealed with high degree of certainty that structure determination of PF2048.1 can be accomplished with as little as {CN, NH} and {NH} from two alignment media respectively. In addition, we believe that more thorough exploration of REDCRAFT's search options, combined with addition of readily available restraints (such as dihedral restraints) can reduce the needed dataset further. This expectation is rooted on the observance of the results from the {3,3} and {2,1} exercises where dihedral restraints were included as part of REDCRAFT's analysis. Inclusion of dihedral restraint not only helped to recover the structure of PF2048.1, but it produced the most accurate structure (to within 1.001Å of the target protein in the case of {3,3}).

Of notable interest is the anomalous nature of structure determination from the set {3,3} compared to that of {3,1}. The refractory nature of this dataset is peculiar and in contradiction with the principles of information theory. In principle, inclusion of data should not harm the outcome unless the included data introduces a level of noise that is nonuniform and more corrupt in nature than the remainder of the data. There is however the possibility of existing inherent degeneracies from the aforementioned set of vectors that when combined with the heuristics of REDCRAFT, produce the observed anomalies. Our future work will investigate these two conjectures.

Our future investigation is to determine the solution state structure of the protein PF2048.1 from experimental data. Our approach will leverage the conclusions of this work in order to acquire the least amount of data compared to the traditional approach of acquiring the most complete dataset. We are confident that our new approach will reduce the quantity of acquired data by nearly 90% and therefore result in significant reduction in financial and temporal cost of protein structure determination by NMR spectroscopy. Although base on the results reported here, structure determination should be plausible with {CN, NH} & {NH} datasets, our experimental investigation of this protein will proceed based on acquisition of the {CN, NH, CH} & {NH} as preparation for missing and noisy data.

## 5 Acknowledgements

This work was supported by NIH Grant Numbers 1R01GM081793 and P20 RR-016461 to Dr. Homayoun Valafar.